\documentclass[aps,twocolumn,superscriptaddress]{revtex4}
\usepackage{graphicx}

\newcommand{\bra}[1] {\left\langle #1 \right|}
\newcommand{\ket}[1] {\left| #1 \right\rangle}

\begin{document}

\title{Computational and physical complexity of synthesizing random multi-qudit quantum states and unitary operators}

\author{Sahel Ashhab}
\affiliation{Advanced ICT Research Institute, National Institute of Information and Communications Technology, 4-2-1 Nukui-Kitamachi, Koganei, Tokyo 184-8795, Japan}
\affiliation{Research Institute for Science and Technology, Tokyo University of Science, 1-3 Kagurazaka, Shinjuku-ku, Tokyo 162-8601, Japan}
\affiliation{Department of Physics, The University of Tokyo, Tokyo 113-0033, Japan}

\author{Bora Basyildiz}
\affiliation{Institute for Quantum Information and Matter and Department of Physics, California Institute of Technology, Pasadena, CA 91125, USA}

\date{\today}

\begin{abstract}
We analyze the complexity of synthesizing random states and unitary operators in a multi-qudit system in two paradigms. In one case, we consider the situation in which we manipulate the system by applying a sequence of one- and two-qudit quantum gates that constitute the elementary, and universal, gate set. The minimum number of gates required to perform the desired operation represents the computational complexity. In the other case, we consider the situation in which we manipulate the physical system using physical fields with optimized control pulses. The minimum time required to perform the desired operation represents the physical complexity. In both cases, we use analytical arguments in combination with optimal-control-theory numerical calculations to determine the complexity of random operations. We show analytically that the computational complexity of synthesizing random states or unitary operators scales exponentially with the number of qudits. While our numerical results are limited to small systems and are insufficient to reveal the asymptotic scaling laws, they suggest that the physical complexity of synthesizing random quantum states and unitary operators scales more slowly than the computational complexity. We discuss various implications of our results, especially concerning the relationship between random and pseudorandom states and unitary operators.
\end{abstract}

\maketitle

\section{Introduction}
\label{Sec:Introduction}

Quantum computing technology is rapidly advancing on a variety of physical platforms \cite{Ladd,Buluta}. In particular, various milestones along the path to practical applications of quantum computers in superconducting circuits \cite{Kjaergaard} have been achieved, including the demonstration of computational power that rivals or exceeds that of present-day supercomputers \cite{Arute,Kim} and quantum error correction \cite{Sivak,Acharya,Brock}.

Random quantum states and unitary operators have played an important role in the recent progress in quantum computing, e.g.~in the demonstration of quantum advantage \cite{Arute}, and they promise to have additional applications in future quantum information technologies \cite{Onorati}, including cryptography \cite{Saini,Ananth}, information scrambling \cite{Brown}, process benchmarking \cite{Emerson} and quantum algorithms \cite{Sen}.

An important question in the study of random quantum states (RSs) and random unitary operators (RUs) is the complexity, or in other words, the resources needed to synthesize these states and operators. In qubit-based systems, it is well known that synthesizing RSs and RUs requires a number of elementary gates that scales exponentially with the number of qubits \cite{Knill}. A related question is how well one can prepare approximate ensembles with a much lower computational cost \cite{Nakata,Guo,Schuster,Foxman,Cui}. These low-cost ensembles of RSs and RUs are called pseudorandom quantum states (PRSs) and pseudorandom unitary operators (PRUs), respectively \cite{Ananth,Ji}. While PRSs and PRUs are formally defined based on their indistinguishability from RSs and RUs, an important distinction between the different ensembles pertains to their complexities. Remarkably, even though an exponential computational cost is generally required to synthesize RSs and RUs, the computational cost of synthesizing PRSs and PRUs is at most polynomial \cite{Schuster}. In this work, to distinguish RSs and RUs from PRSs and PRUs, we will generally use the terms RSs and RUs to refer to those states and unitary operators that require an exponential number of gates to synthesize. In fact, this property applies to all states and unitary operators, except for measure-zero subsets that have polynomial or sub-polynomial cost.

Although much of quantum information research is based on binary logic using qubits, the use of $d$-level quantum systems, or qudits, has been gaining an increasing amount of attention in recent years \cite{Wang}. Some of the advantages of using qudits include the increased information capacity of the hardware, increased noise resilience of quantum error correction protocols and increased security of quantum cryptography protocols \cite{Wang,Campbell2014,Islam2017}. In the past few years, several experiments have demonstrated a number of milestones in the physical relization of qudits, including in superconducting circuits \cite{Neeley,Yurtalan,Blok,Morvan,Kononenko,Goss,Tripathi,Champion}, trapped ions \cite{Senko,Ringbauer,Nikolaeva,Low,Shi}, defect nuclear spins in silicon \cite{FernandezDeFuentes,Yu}, arrays of Rydberg atoms \cite{Robert2025,Burshtein2025},  and photonic circuits \cite{Kues,Chi}.

As with qubits, it is crucial for the optimal utilization of qudits to understand the scaling of the resources needed to synthesize their quantum states and unitary operators. Recently, we investigated the minimum times needed to implement qutrit (3-level system) and ququart (4-level system) controlled-Z (CZ) gates in superconducting circuits and found that, rather surprisingly, they are essentially the same as the minimum gate time for the qubit CZ gate \cite{Basyildiz2025}. This result sets a benchmark for future studies on qudit gates. As another important reference point, the resources needed to synthesize multi-qudit RSs and RUs can guide future developments in the field of qudit-based quantum information processing.

In this work, we investigate the complexity of synthesizing multi-qudit RSs and RUs. There are two different paradigms in the study of synthesizing quantum states and unitary operators. In the first, one uses a sequence of elementary quantum gates to implement the desired operation. In analogy to previous work on qubits \cite{ShendeUnitary,Plesch}, we derive mathematical lower bounds for the number of gates needed to implement various operations with qudits. We perform numerical calculations for small systems to develop some understanding of how tight these lower bounds are. This approach defines the computational complexity of quantum operations. The second paradigm in the study of synthesizing quantum states and unitary operators is that based on the evolution of the physical system under Hamiltonian dynamics with appropriate control pulses. Using optimal control theory (OCT) calculations, we determine the minimum times needed to synthesize RSs and RUs. This approach defines the physical complexity of quantum operations. An intriguing question here is what role the continuous-time nature of Hamiltonian dynamics plays in resource quantification. Specifically, an important question is whether an exponentially long time is needed to synthesize RSs and RUs or whether the infinite amount of information contained in a continuous-time control pulse is sufficient to provide the ability to access all parts of the target space in a sub-exponential evolution time. This question is motivated by the fact that such parameter counting considerations are a crucial ingredient in derivations of the lower bounds for the number of elementary gates needed to synthesize RSs and RUs.

The remainder of this paper is organized as follows: In Sec.~\ref{Sec:ComputationalComplexity}, we investigate the computational complexity of RSs and RUs, including the derivation of mathematical lower bounds and numerical calculations of gate counts. In Sec.~\ref{Sec:PhysicalComplexity}, we investigate the physical complexity of the same tasks by numerically evaluating the minimum times needed to perform these tasks. In Sec.~\ref{Sec:Discussion}, we discuss a few implications of our results. One of the main questions pertains to the relation between RSs and PRSs, and similarly the relation between RUs vs PRUs. We conclude with some final remarks in Sec.~\ref{Sec:Conclusion}.

\section{Computational complexity}
\label{Sec:ComputationalComplexity}

In this section, we investigate the problem of synthesizing an RS or RU using a sequence of elementary quantum gates. These gate sequences are often called quantum circuits. We consider quantum circuits based on two-qudit gates as entangling gates. In other words, the quantum circuit is a sequence of single-qudit and two-qudit gates. As is common in complexity analysis of quantum circuits, we assume that single-qudit gates are easy to implement, with high fidelity and short gate time. As such, they contribute a negligible cost in terms of the required resources, e.g.~implementation time \cite{Ladd,Buluta,Vidal}. Implementing the entangling gates therefore represents the main contribution to the computational cost. Specifically, the two-qudit gate count is the key metric for evaluating the cost. We refer to the number of two-qudit gates as the circuit size. It is worth noting that in some of the leading quantum computing architectures, such as superconducting qudits with weak anharmonicities, the single-qudit and two-qudit gate times are of the same order of magnitude. As such, the time needed to implement single-qudit gates can be non-negligible. Nevertheless, since the number of single-qudit gates in the quantum circuit is of the same order as the number of entangling gates, the entangling gate count is still a good metric to quantify quantum circuit complexity.

\subsection{Lower bounds}

We start by quickly reviewing the arguments and derivation of the lower bounds for synthesizing an $n$-qubit RS or RU using the two-qubit CZ gate as the entangling gate \cite{ShendeUnitary,Plesch}. The derivation involves counting the number of adjustable parameters in the quantum circuit and comparing this number with the number of independent parameters in the target state or operator. The latter can also be thought of as the number of dimensions of the space of all possible target states or operators. We consider a quantum circuit that contains $N$ CZ gates. If we take into consideration the fact that any sequence of single-qubit gates applied to a qubit is equivalent to one single-qubit gate, we find that the $N$ CZ gates will be surrounded by at most $n+2N$ non-redundant single-qubit gates. These are broken down as follows: We can apply $n$ single-qubit gates to the $n$ qubits at the beginning of the quantum circuit. Then, after each CZ gate, we can apply single-qubit gates to the two qubits involved in the CZ gate. Any additional single-qubit gates are redundant and can be merged with other single-qubit gates that appear earlier in the quantum circuit. In general, a single-qubit gate is defined by three parameters, suggesting that there are $3(n+2N)$ adjustable parameters in the quantum circuit. However, it turns out that this number can be reduced. A single-qubit gate can be decomposed into a rotation about the z axis followed or preceded by a rotation about an axis in the xy plane. Since z-axis rotations commute with the CZ gate, many of the z-axis rotations can be eliminated from the quantum circuit without affecting the set of states or unitary operators that can be synthesized using the quantum circuit. In other words, many z-axis rotations are redundant. In the case of state preparation, all z-axis rotations can be eliminated, which gives a total of $2n+4N$ adjustable parameters in a quantum circuit of $N$ CZ gates. In the case of unitary operator synthesis, the first single-qubit gate applied to each qubit is a general three-dimensional operation, whereas all subsequent single-qubit rotations can be treated as two-dimensional operations, which gives a total of $3n+4N$ adjustable parameters. Having counted the number of non-redundant adjustable parameters, we now need the number of independent parameters in an arbitrary target. For state preparation, the number of independent parameters in a general $n$-qubit state is $2\times 2^{n}-2$. For unitary operators, the number is $4^n-1$. The lower bounds for the number of CZ gates needed to synthesize an arbitrary target are then given by $N=\lceil (2^{n}-1-n)/2 \rceil$ for state preparation and $N=\lceil (4^n-1-3n)/4 \rceil$ for unitary operator synthesis, where $\lceil x \rceil$ denotes the ceiling function, i.e.~the smallest integer larger than $x$. This number of CZ gates ensures that the quantum circuit has enough adjustable parameters to match or exceed the number of free parameters in an arbitrary target.

We can now generalize the above derivation to the case of synthesizing an $n$-qudit quantum state or unitary operator using a sequence of single-qudit gates and two-qudit entangling gates. For the entangling gate, we focus on the two-qudit CZ gate \cite{Goss}, which is known to be a universal gate for multi-qudit systems:
\begin{equation}
\widehat{CZ} = \sum_{k,l=0}^{d-1} e^{i2\pi k l /d} \ket{k,l} \bra{k,l}.
\end{equation}
Here, the indices $k$ and $l$ are the quantum states of the two coupled qudits. A general quantum state of $n$ qudits has $d^n$ complex basis state amplitudes. Since these amplitudes must obey the normalization constraint, and the overall phase of a quantum state is physically irrelevant, we find that the number of physically meaningful free parameters that specify the state is $2\times d^n - 2$. In other words, the space of $n$-qudit quantum states has dimension $2\times d^n - 2$. A general unitary operator of $n$ qudits can be represented by a $d^n \times d^n$ unitary matrix, which contains $d^{2n}$ complex matrix elements, each with a real and an imaginary part. After considering the $d^{2n}$ orthonormality constraints and eliminating one irrelevant overall phase factor, we find that an $n$-qudit unitary operator is specified by $d^{2n} - 1$ independent parameters.

As in the case of qubits, the number of non-redundant single-qudit gates in a quantum circuit that contains $N$ entangling gates is $n+2N$. The most general single-qudit unitary operator has $d^2$ independent parameters. If we eliminate one overall-phase parameter, we have $d^2-1$ independent parameters. In cases where we can eliminate the z-axis rotations, a single-qudit unitary operator has $d^2-d$ independent parameters. This point was discussed in detail in Refs.~\cite{Yurtalan,Kononenko,Iliat} for the case of qutrits. If we consider these parameter-count changes in comparison with the qubit case, we find that the number of independent parameters in an $n$-qudit-$N$-entangling-gate quantum circuit is $d(d-1)(n+2N)$ for state preparation and $(d^2-1)n+2d(d-1)N$ for unitary operator synthesis.

The lower bounds for the number of two-qudit CZ gates required to synthesize arbitrary quantum states and unitary operators are then given by
\begin{equation}
N_{S} = \left\lceil \frac{2\times d^n - 2 - d(d-1)n}{2d(d-1)} \right\rceil
\label{Eq:LowerBoundState}
\end{equation}
for state preparation and
\begin{equation}
N_{U} = \left\lceil \frac{d^{2n} - 1 - (d^2-1)n}{2d(d-1)} \right\rceil
\label{Eq:LowerBoundUnitary}
\end{equation}
for unitary operator synthesis. Examples of gate count lower bounds for a few different combinations of $n$ and $d$ values are shown in Table \ref{Table:MinimumGateCount}. The quantum circuit size lower bounds clearly scale exponentially with the number of qudits. Specifically, if we consider the large $n$ and $d$ limits, we obtain the asymptotic scaling laws
\begin{eqnarray}
N_{S} & \sim & d^{n-2},
\label{Eq:LowerBoundStateAsymptotic}
\\
N_{U} & \sim & \frac{d^{2n-2}}{2}.
\label{Eq:LowerBoundUnitaryAsymptotic}
\end{eqnarray}

\begin{center}
\begin{table}
\scriptsize
\begin{tabular}{ || c | c || c | c || c | c || }
\hline
\multicolumn{2}{||c||}{} & \multicolumn{2}{|c||}{State preparation} & \multicolumn{2}{|c||}{Unitary operator synthesis} \\
\hline
$n$ & $d$ & Lower bound & Numerical & Lower bound & Numerical \\
\hline
\hline
2 & 2 & 1 & 1 & 3 & 3 \\
\hline
3 & 2 & 2 & 3 & 14 & 14 \\
\hline
4 & 2 & 6 & 6 & 61 & $\leq$ 62 \\
\hline
\hline
2 & 3 & 1 & 1 & 6 & 6 \\
\hline
3 & 3 & 3 & 4 & 59 & \\
\hline
4 & 3 & 12 & $\leq$ 18 & 544 & \\
\hline
\hline
2 & 4 & 1 & 1 & 10 & 10 \\
\hline
3 & 4 & 4 & 6 & 169 & \\
\hline
4 & 4 & 20 & $\leq$ 32 & 2729 & \\
\hline
\end{tabular}
\caption{Number of two-qudit CZ gates $N$ needed to synthesize an arbitrary quantum state ($N_S$) or unitary operator ($N_U$) for a few combinations of number of qudits $n$ and qudit dimension $d$. The lower bound values are based on Eqs.~(\ref{Eq:LowerBoundState}) and (\ref{Eq:LowerBoundUnitary}). The numerical results were obtained using the exhaustive search approach, whenever possible. Values given with the $\leq$ sign indicate cases where we used the probabilistic approach, which gives an upper bound for the minimum required number of entangling gates $N$. Blank fields correspond to cases that are computationally infeasible for our calculations to tackle.}
\label{Table:MinimumGateCount}
\end{table}
\end{center}

We finally note that the gate counts derived in this subsection are mathematical lower bounds for arbitrary targets. They are not guaranteed to be tight. Indeed, we will see shortly that they are sometimes but not always tight, and there is no obvious rule that we can use to predict when they will be tight. On the other hand, there are specific targets that can be synthesized with fewer gates than the above lower bounds. A well-known example is the multi-qubit CZ gate.

\subsection{Numerical evaluation of computational complexity}

We performed numerical calculations to determine the number of entangling gates needed to synthesize RSs and RUs for a few combinations of $n$ and $d$ values. The calculations are generalized versions of those performed in Refs.~\cite{Ashhab2022CircuitSynthesis,Ashhab2024} for qubits. When the number of qudits and the number of entangling gates are small, we can perform an exhaustive search where we consider all gate configurations \cite{Ashhab2022CircuitSynthesis}. For situations with more than very few qudits and gates, where the exhaustive search approach is infeasible, we use a probabilistic search \cite{Ashhab2024}, which turns out to be a powerful and efficient approach to find optimal or near-optimal quantum circuits. In the present work, the target RSs and RUs are generated using the QR-decomposition-based method for generating Haar-random quantum states and unitary operators \cite{Stewart}.

A gate configuration is defined by the pairs of qudits coupled in each entangling gate in the quantum circuit. As mentioned above, for the entangling gate, we use the two-qudit CZ gate. A single entangling gate has $n(n-1)/2$ possible configurations, i.e.~possible pairs of qudits that are coupled by the gate. A sequence of $N$ entangling gates has $[n(n-1)/2]^N$ possible configurations. For each configuration, we allow the single-qudit unitary operators in the quantum circuit to be variable operators, and we optimize their parameters to maximize the fidelity. If $N$ is below a certain threshold, the maximum achievable fidelity is below 1. We gradually increase $N$ and perform the optimization calculations until we identify the smallest value of $N$ that allows us to achieve perfect synthesis with 100\% fidelity. In practice, the fidelity is limited by numerical errors, such that we do not expect to obtain exactly $F=1$. However, the typical situation that we encounter in our calculations is that a certain number of entangling gates $N-1$ gives fidelities of a few nines, i.e.~close to but clearly different from 1, while adding one entangling gate for a total of $N$ entangling gates gives $F=1$ to within numerical errors, which are typically on the order of $10^{-12}$ in our calculations. It is therefore often easy to identify the minimum number of gates required to achieve perfect state or unitary operator synthesis.

In the probabilistic approach, instead of analyzing each one of the exponentially large number of configurations, we try 100 randomly generated quantum circuits. If we find at least one quantum circuit that gives $F=1$ for $N$ entangling gates, we can say with certainty that the minimum number required to synthesize the target is at most $N$. As shown in Ref.~\cite{Ashhab2024}, this probabilistic approach is an efficient approach to find surprisingly tight upper bounds on the number of entangling gates required for perfect synthesis. For larger systems, where the number of qudits $n$ is large, the number of gates $N$ is expected to grow exponentially with $n$, as can be seen in Eqs.~(\ref{Eq:LowerBoundState}) and (\ref{Eq:LowerBoundUnitary}). As a result, even the probabilistic approach quickly becomes infeasible, because analyzing even a single configuration becomes intractable.

The results of our numerical calculations are shown in Table \ref{Table:MinimumGateCount}. In cases where we performed an exhaustive search, we simply provide the number of entangling gates, being confident that it is the exact number required for perfect synthesis. In cases where we used the probabilistic approach, we use the $\leq$ sign, because we cannot exclude the possibility that perfect synthesis can be performed with fewer gates, unless the numerically obtained value of $N$ matches the mathematical lower bound. In cases where even the probabilistic approach is infeasible, we leave the box blank.

The numerical results are consistently close to the lower bounds. This result indicates that the lower bounds are generally tight, although not always exactly tight. This result follows a similar result for qubits \cite{Ashhab2022CircuitSynthesis,Ashhab2024}. Table \ref{Table:MinimumGateCount} also shows that the computational cost grows rapidly. In particular, for unitary operator synthesis, we were able to treat only five cases before the computational cost became prohibitive for our desktop computers.

It is worth mentioning here that the CZ gate is not optimal for minimizing the gate count. In the case of qubits, the so-called $B$ gate is optimal \cite{Zhang,Ashhab2024}. The reason why the $B$ gate gives lower gate counts compared to the CZ gate is that the $B$ gate does not commute with any single-qubit gate, which leads to a larger number of adjustable parameters for the same gate configuration. However, for practical purposes, the CZ gate is easier to implement than the $B$ gate, making the CZ gate a better choice as the entangling gate in a realistic experimental setup \cite{Ashhab2024}. In the case of qudits, we can similarly perform a gate count calculation for the case when the two-qudit entangling gate does not commute with any single-qudit gates. In this case, the single-qudit gates that follow entangling gates will have $d^2-1$ parameters each, instead of the $d^2-d$ in the case of the CZ gate. The number $d^2-1$ is the total number of generalized Gell-Mann matrices for a qudit of dimension $d$, excluding the one proportional to the identity. The lower bounds for the entangling gate count then become
\begin{equation}
N_{S} = \left\lceil \frac{2\times d^n - 2 - d(d-1)n}{2(d^2-1)} \right\rceil
\end{equation}
for state preparation and
\begin{equation}
N_{U} = \left\lceil \frac{d^{2n} - 1 - (d^2-1)n}{2(d^2-1)} \right\rceil
\end{equation}
for unitary operator synthesis. These lower bounds are, in general, lower than those for the CZ gate. Note that the difference between these formulae and the ones in Eqs.~(\ref{Eq:LowerBoundState}) and (\ref{Eq:LowerBoundUnitary}) is roughly a factor of $(d^2-1)/(d^2-d)$. In other words, the exponential scaling is not affected by using a gate-count-optimal entangling gate. Furthermore, as mentioned above, there are known protocols for implementing the two-qudit CZ gate \cite{Goss}. If we want to use an alternative entangling gate, we must identify a gate that minimizes the gate count and determine how to implement it using the physical system Hamiltonian. If there is no simple way to implement the gate, this supposedly optimal gate could be less optimal for implementing quantum operations in practice, similarly to the case of the two-qubit CZ and $B$ gates \cite{Ashhab2024}. We will therefore not perform any further calculations on alternative entangling gates other than the two-qudit CZ gate.

\section{Physical complexity}
\label{Sec:PhysicalComplexity}

The second approach to synthesizing quantum states and unitary operators is to apply physical control signals via external fields, e.g.~electromagnetic fields, that couple to some dynamical variables of the system encoding the qudits. The system then evolves according to the Schr\"odinger equation. With the proper control pulses, the desired target state or unitary operator is synthesized.

In a typical physical setting, the Hamiltonian is given by
\begin{equation}
\hat{H} = \hat{H}_0 + \sum_{j=1}^{M} f_j(t) \hat{H}_j,
\label{Eq:ControlHamiltonianGeneral}
\end{equation}
where $\hat{H}_0$ is the fixed part of the Hamiltonian, while $\hat{H}_j$ are the control operators, and $f_j(t)$ are the control pulses. The control pulses can be optimized using OCT techniques to maximize the fidelity between the dynamically obtained quantum state or unitary operator and the target.

If the total pulse time is not sufficiently long, the fidelity will be below 1, even for the optimal pulse. Conversely, if the pulse time is sufficiently long, it is at least theoretically possible to achieve fidelity $F=1$, i.e.~perfect synthesis. An underlying implicit assumption here is that the system is sufficiently controllable. In other words, the Hamiltonian in Eq.~(\ref{Eq:ControlHamiltonianGeneral}) has enough control capability to allow reaching the target. The conditions of controllability are rather easy to satisfy. In the case of a qubit array, if at least one single-qubit control term per qubit is available, and there are enough two-qubit interactions to create a (possibly minimally) connected graph, the system is typically fully controllable.

As a typical case that applies to some of the common superconducting qudit architectures, we use the $n$-qudit Hamiltonian
\begin{equation}
\hat{H} = \hat{H}_{\rm sq,fixed} + \hat{H}_{\rm sq,control} + \hat{H}_{\rm int},
\label{Eq:ControlHamiltonianQudits}
\end{equation}
where
\begin{eqnarray}
\hat{H}_{\rm sq,fixed} & = & \sum_{k=1}^{n} \sum_{j=0}^{d-1} \left( \omega_k j + \eta_k \frac{j(j-1)}{2} \right)  \hat{\Pi}_{k,j}, \nonumber \\
\hat{H}_{\rm sq,control} & = & \sum_{k=1}^{n} \sum_{j=0}^{d-2} \sum_{l=j+1}^{d-1} f_{k,j,l}(t) \left( \ket{j}_k \bra{l}_k + \ket{l}_k\bra{j}_k \right), \nonumber \\
\hat{H}_{\rm int} & = & \sum_{k=1}^{n-1} \sum_{m=k+1}^{n} g \left( \hat{a}_k + \hat{a}_k^{\dagger} \right) \otimes \left( \hat{a}_m + \hat{a}_m^{\dagger} \right),
\end{eqnarray}
$\omega_k$ are the different qudit frequencies, $\eta_k$ are the qudit anharmonicities, $f_{k,j,l}(t)$ are the control pulses coupling all the possible single-qudit energy-level pairs, $g$ is the inter-qudit coupling strength, $\hat{\Pi}_{k,j}$ is the projection operator of qudit $k$ on state $\ket{j}$, i.e.~$\hat{\Pi}_{k,j}=\ket{j}_k \bra{j}_k$, the operators $\hat{a}_k$ and $\hat{a}^{\dagger}_k$ are, respectively, the truncated annihilation and creation operators
\begin{eqnarray}
\hat{a}_k & = & \sum_{j=1}^{d-1} \sqrt{j} \ket{j-1}_k \bra{j}_k, \nonumber \\
\hat{a}_k^{\dagger} & = & \sum_{j=1}^{d-1} \sqrt{j} \ket{j}_k \bra{j-1}_k.
\label{Eq:LadderOperators}
\end{eqnarray}
The factor $\sqrt{j}$ is the usual factor in the creation and annihilation operators of a harmonic oscillator. These operators are indeed good approximations for some of the standard superconducting qubit designs, such as the transmon, with the caveat that one must always be careful about leakage to quantum states outside the truncated harmonic oscillator state space.

A few points are worth emphasizing about the form of the Hamiltonian that we use in this work. First, we assume that all transitions (i.e.~$\ket{j} \leftrightarrow \ket{l}$) in a qudit can be controllably and independently driven. It is common in superconducting qudit circuits that the same driving field couples to all transitions and that other circuit properties limit the control field's power and bandwidth. These limitations often slow down single-qudit operations. While our optimistic assumptions are not applicable to typical experimental situations, they are convenient for our theoretical study, since in theoretical studies on composite-system speed limits the interaction terms represent the bottleneck for the operation speed. Hence it is preferable to avoid adding the complication of restricting the single-qudit control terms. Second, the interaction Hamiltonian $\hat{H}_{\rm int}$ describes a situation with all-to-all interactions, i.e.~a fully connected graph. Here also we are making an experimentally optimistic assumption to ensure that we capture the most fundamental speed limit. Using a Hamiltonian with restricted connectivity, such as nearest-neighbor interactions, would lead to longer operation times in general. Such situation are in fact the standard for superconducting quantum processors. It is worth noting that the above assumptions about full single-qudit controllability and all-to-all connectivity are consistent with similar ones that we made when analyzing the computational complexity in Sec.~\ref{Sec:ComputationalComplexity}. These assumptions are therefore appropriate for the purpose of making a fair comparison between the two complexity measures. Third, since in this work we do not impose constraints on the strength of the control fields, the OCT calculation can be simplified by ignoring the term $\hat{H}_{\rm sq,fixed}$ in the Hamiltonian, which allows us to reduce the number of time steps while satisfying the condition that the time step is the smallest time scale in the calculation. The minimum synthesis time with this simplification will be the same as that of the original problem \cite{Basyildiz2023}.

A natural time scale to use as a reference value in this study is the minimum time needed to perform a CZ gate in a two-qubit system. This time is given by $T_{CZ2}=\pi/(4g)$. As was shown in Ref.~\cite{Basyildiz2025}, this time is also the qutrit and ququart CZ gate time in typical superconducting qudit designs. We will use this time as the time unit when we report our simulation results below.

We determine the minimum operation time of each specific case using numerical OCT calculations. The optimization calculations are similar to those of Refs.~\cite{Ashhab2012,Basyildiz2025}, which used the gradient ascent pulse engineering (GRAPE) algorithm \cite{Khaneja}. We start with a short pulse time and gradually increase it until the fidelity approaches $F=1$. We identify the minimum gate time as the time at which the fidelity crosses the threshold $F=0.999$. We note that the numerically obtained gate time is rather insensitive to the exact threshold value. We expect that other error sources, e.g.~the fact that we choose a spacing of $\sim 10 \%$ between gate time values and incomplete convergence of the optimization procedure, set error bars of $\sim 10 \%$ in our results for the minimum synthesis time. This error is sufficiently small for our main purpose, which is to distinguish between different scaling behaviors of the complexity, e.g.~sublinear versus polynomial versus exponential. For each combination of $n$ and $d$ values we use five different targets chosen randomly from a Haar-uniform distribution. For each target we run the algorithm five times with different random seeds for the control pulse, and we take the highest obtained fidelity from the five runs. The total pulse time is divided into $10^3$ time steps and the algorithm is run for $10^4$ optimization iterations.

The calculation results are shown in Table \ref{Table:MinimumGateTime}. The OCT calculation time limits us to only small values of $n$ and $d$. The few combinations that we can treat numerically allow us to see some preliminary trends in the minimum synthesis time as a function of $n$ and $d$, but we cannot infer the asymptotic scaling laws. Nevertheless, the few trends that we can see in the numerical data contain some useful insights. For example, the minimum times for state preparation and unitary operator synthesis are drastically different from each other and from those that we obtained for the computational complexity.

\begin{center}
\begin{table}
\scriptsize
\begin{tabular}{ || c | c || c || c || }
\hline
$n$ & $d$ & State preparation & Unitary operator synthesis \\
\hline
\hline
2 & 2 & 0.2, 0.2, 0.6, 0.46, 0.44 (0.02) & 1.7, 1.8, 1.5, 1.8, 1.4 (0.1) \\
\hline
3 & 2 & 0.56, 0.54, 0.54, 0.52, 0.62 (0.02) & 3.2, 3.2, 3.2, 3.2, 3 (0.2) \\
\hline
4 & 2 & 0.9, 0.95, 0.85, 0.7, 0.85 (0.05) & 8.5, 8.5, 8.5, 8.5, 9 (0.5) \\
\hline
\hline
2 & 3 & 0.26, 0.32, 0.26, 0.36, 0.22 (0.02) & 1.5, 1.6, 1.6, 1.7, 1.7 (0.1) \\
\hline
3 & 3 & 0.5, 0.5, 0.46, 0.5, 0.54 (0.02) & 7, 7, 7, 7, 7 (0.5) \\
\hline
4 & 3 & & \\
\hline
\hline
2 & 4 & 0.26, 0.2, 0.26, 0.24, 0.28 (0.02) & 1.8, 1.8, 1.8, 1.8, 1.8 (0.1) \\
\hline
3 & 4 & & \\
\hline
4 & 4 & & \\
\hline
\end{tabular}
\caption{Minimum time needed to synthesize $n$-qudit random targets for a few combinations of number of qudits $n$ and qudit dimension $d$. All the results in this table are in units of the two-qubit CZ gate time $T_{CZ2}$. In each set, we performed five calculations for five randomly generated targets. The gate times for the five targets in each set are shown. The number in parentheses shows the distance between pulse time values in the calculations and therefore serves as an estimate for the uncertainty in the calculated time.}
\label{Table:MinimumGateTime}
\end{table}
\end{center}

In the case of state preparation, while the operation time grows as a function of $n$, the growth is rather slow. For example, if we compare the cases of two- and three-qubit state preparation, the instance-to-instance variations in the operation time are larger than the difference between the average times in the two cases. The slowness in the overall growth can be partly explained by the all-to-all connectivity in our model, which means that the quantum state can spread among all the different qudits in parallel, not between pairs of qudits one at a time. Another result that we note here is that the growth of the operation time as a function of $n$ is slower than that shown in Table \ref{Table:MinimumGateCount} for the computational complexity. If we focus on the qudit dimension $d$, we can see that the operation time is only weakly dependent on $d$. In fact, it seems that the operation time decreases with increasing $d$. Two factors are likely to contribute to this seemingly counter-intuitive result. One factor is the increasing inter-qudit coupling strength as we go up the energy level ladder, and the other factor is constructive interference enabled by the existence of multiple paths between quantum states in the bi-linear interaction model. This result is not completely surprising, considering similar results obtained in Refs.~\cite{Ashhab2022SpeedLimits,Basyildiz2023}.

In the case of unitary operator synthesis, we see more pronounced changes in the operation time as we vary $n$ and $d$. For example, if we compare the three cases $n=2,3,4$ with $d=2$, the operation time increases by a factor between 2 and 3 every time we increase $n$ by 1. Furthermore, for qutrits ($d=3$), when we increase $n$ from 2 to 3, the operation time increases by a factor of about 4. This rather strong dependence suggests that the scaling could be exponential. Still, the increase is slower than that shown in Table \ref{Table:MinimumGateCount} for the computational complexity. As in the case of state preparation, if we compare the three cases with two-qudits ($n=2$ and $d=2,3,4$), we can see that the operation time is weakly dependent on $d$. However, if we take $n=3$, we see a clear increase in the operation time when we go from $d=2$ to $d=3$. Again, the increase is smaller than that shown in Table \ref{Table:MinimumGateCount}.

In all cases represented in Table \ref{Table:MinimumGateTime}, with the exception of state preparation with $n=2$ and $d=2$, the instance-to-instance variations are relatively small. In general, the variations seem to become smaller as $n$ and $d$ increase, especially for unitary operator synthesis. The instance-to-instance variations in Table \ref{Table:MinimumGateTime} give us an idea about the space of quantum states and unitary operators, in particular the distance between general targets and the origin, which is defined by the trivial operation of taking zero evolution time or, alternatively, not applying any gates.

\section{Discussion}
\label{Sec:Discussion}

\subsection{Random versus pseudorandom ensembles}

Our results in Sections \ref{Sec:ComputationalComplexity} and \ref{Sec:PhysicalComplexity} have implications for the relationship between RUs and PRUs, and similarly between RSs and PRSs. We discuss a few relevant points here. For definiteness, we focus on RUs and PRUs.

As explained in Sec.~\ref{Sec:ComputationalComplexity}, synthesizing an RU generally requires a number of elementary gates that grows exponentially with the number of qudits \cite{RandomUnitaryFootnote}. In contrast, PRUs represent a subset of unitary operators that require gate counts that grow at most polynomially with system size, while still serving as ensembles that accurately reproduce the statistical properties of RUs under any probe protocol that uses at most a polynomial number of one- and two-qudit gates. It is worth mentioning here that PRUs represent a subset of measure zero in the set of all unitary operators. This situation naturally raises a few questions.

One question relates to the physical complexity of PRUs. If PRUs can be synthesized with a polynomial number of elementary gates, this scaling law will set an upper bound on the time needed to synthesize PRUs in the continuous-time approach, since we can always turn a gate sequence into a control pulse that implements the gates one by one. This gate-to-pulse conversion would give the same scaling law for the discrete- and continuous-time approaches. The continuous-time approach could give faster total operation times, because it is possible that there can be optimized pulses that implement the full target operation faster than the step-by-step pulse that strictly follows the optimal gate sequence taken from the discrete-time approach. We can therefore say that the physical complexity of synthesizing PRUs scales at most polynomially with the number of qudits.

Another important point in this context is that in the vicinity of any PRU, there must exist exponential-computational-complexity unitary operators that are arbitrarily close to the PRU. To see this point, we can consider a PRU $\hat{U}_{\rm PR}$. Now we take a generator $\hat{G}_R$, i.e.~a $d^n \times d^n$ Hermitian matrix, that generally contains finite, random and uncorrelated matrix elements throughout the matrix, including all off-diagonal matrix elements. To be consistent with the convention used in previous sections, we just exclude any component proportional to the identity matrix and hence take $\hat{G}_R$ to be traceless. We then take the product
\begin{equation}
\hat{U}_{\rm R} = \exp \left\{ i \gamma \hat{G}_R \right\} \hat{U}_{\rm PR},
\label{Eq:RUfromPRU}
\end{equation}
where $\gamma$ is a real number that we will treat as a small variable parameter. The operator $\hat{U}_{\rm R}$ can be thought of as a new unitary operator that is obtained by slightly deforming $U_{\rm PR}$. Since the generator $\gamma\hat{G}_R$ contains $d^{2n}-1$ independent variables, the space of unitary operators defined by this procedure must have $d^{2n}-1$ dimensions. This number of dimensions is the same as that of the whole set of unitary operators. Most of the unitary operators defined by Eq.~(\ref{Eq:RUfromPRU}) must therefore belong to the most general set of unitary operators, which require an exponential number of gates to synthesize. They are therefore not PRUs. This statement remains true even if we take $\gamma$ to be infinitesimally small. When $\gamma$ is infinitesimally small, all the unitary operators in the set are infinitesimally close to $\hat{U}_{\rm PR}$ by any reasonable definition of distance between unitary operators. We can therefore conclude that there must exist (infinitely many) unitary operators with exponential computational complexity that are infinitesimally close to any PRU.

This argument has a remarkable implication: even if we take a pair of unitary operators, one general RU and one PRU, that are extremely close to each other, the RU will have an exponentially large computational complexity, while the PRU will have an at most polynomially large computational complexity.

In contrast to this unusual property of the computational complexity, we would now like to establish that the physical complexity of a PRU and an infinitesimally close unitary operator must be the same. We consider the minimum time needed to implement the operation $\exp \left\{ i \gamma \hat{G}_R \right\}$ given in Eq.~(\ref{Eq:RUfromPRU}). This time ($T_{\gamma G}$) sets bounds for the minimum time ($T_{\rm R}$) needed to synthesize $\hat{U}_{\rm R}$, assuming that we know the minimum time ($T_{\rm PR}$) needed to synthesize $\hat{U}_{\rm PR}$. Specifically, by a generalization of the Cauchy-Schwarz inequality, we find that $T_{\rm PR}-T_{\gamma G} \leq T_{\rm R} \leq T_{\rm PR}+T_{\gamma G}$. As $\gamma$ is infinitesimally small, we can write
\begin{equation}
\exp \left\{ i \gamma \hat{G} \right\} \approx \hat{1} + i \gamma \hat{G}.
\label{Eq:PerturbationApproximation}
\end{equation}
If we set the control Hamiltonian $\hat{H}$ to a time-independent value and let the system evolve for a finite time $\tau$, we obtain the evolution operator
\begin{equation}
\exp \left\{ -i \hat{H} \tau \right\} \approx \hat{1} - i \hat{H} \tau.
\end{equation}
The $M$ adjustable parameters $f_j(t)$ that are associated with the $M$ control operators $\hat{H}_j$ define an $M$-dimensional space of unitary operators that can be generated in time $\tau$, keeping in mind that we are assuming time-independent Hamiltonian settings for the time being. If there is a choice of coefficients $f_j$ for the different terms in $\hat{H}$ that produces a Hamiltonian that is proportional to $\hat{G}$ in Eq.~(\ref{Eq:PerturbationApproximation}), this operation can be implemented in a single step with a time-independent Hamiltonian. Any unitary operator in this space can be synthesized in a time that is linearly proportional to $\gamma$. In other words, if we take one element in the set of unitary operators generated in this way, expressed as $\exp \left\{ i \gamma \hat{G} \right\}$, and cut $\gamma$ in half, the time needed to synthesize the new unitary operator will also be cut in half. If instead of a time-independent Hamiltonian we consider the four step sequence
\begin{equation}
\exp \left\{ i \hat{H}_b \tau \right\} \exp \left\{ i \hat{H}_a \tau \right\} \exp \left\{ -i \hat{H}_b \tau \right\} \exp \left\{ -i \hat{H}_a \tau \right\},
\end{equation}
the Magnus expansion,
\begin{eqnarray}
\exp \Bigg\{ & & -i \int_{0}^{4\tau} \hat{H}(t) dt \nonumber \\
& & - \frac{1}{2} \int_{0}^{4\tau} dt_1 \int_{0}^{t_1} dt_2 [\hat{H}(t_1), \hat{H}(t_2)] + \dots \Bigg\},
\end{eqnarray}
gives the effective unitary evolution operator
\begin{equation}
\exp \left\{ \left[ \hat{H}_a , \hat{H}_b \right] \tau^2 + \mathcal{O}\left(\tau^3\right) \right\}.
\label{Eq:EvolutionOperatorSecondOrder}
\end{equation}
As a result, the commutators between the $M$ different control terms $\hat{H}_j$ in addition to the fixed part of the Hamiltonian in general define $M(M+1)/2$ directions for the evolution of the synthesized unitary operator. Note that these commutators could contain some of the original $M+1$ operators ($M$ control operators and one fixed term) or could define $M(M+1)/2$ additional directions for evolution in the space of unitary operators, depending on the specific details of the situation under study. If we now assume that Eq.~(\ref{Eq:EvolutionOperatorSecondOrder}) generates a unitary operator of the form in Eq.~(\ref{Eq:PerturbationApproximation}), and we cut $\gamma$ in half, we must reduce $\tau$ to $\tau^{1/2}$ to obtain the new unitary operator. If we go to higher orders and take more elaborate pulse sequences, we obtain increasingly complex commutator expressions that eventually account for all $d^{2n}-1$ possible directions necessary to explore the full space of unitary operators in the vicinity of the unperturbed PRU (this statement being guaranteed by our assumption that the system is fully controllable). This exercise produces a hierarchy of evolution directions, with the scaling law $\tau\propto\gamma^{1/k}$, where the integer $k$ depends on the number of commutators that we need to take to reach the respective operator. Let us assume that the ability to access all $d^{2n}-1$ evolution directions is achieved after taking $K$ commutators between the control operators $\hat{H}_j$ and the fixed part of the Hamiltonian. In this case, generating the operator $\exp \left\{ i \gamma \hat{G} \right\}$ will take a time $\tau$ that scales at most as $\gamma^{1/K}$. In the limit $\gamma\rightarrow 0$, we obtain $\tau\rightarrow 0$, regardless of the exact value of $K$, which means that the unitary operator $\exp \left\{ i \gamma \hat{G} \right\}$ with an infinitesimally small value of $\gamma$ can be synthesized in an infinitesimally short duration. Therefore, if we can synthesize the PRU $\hat{U}_{\rm PR}$ in time $T$, we can be certain that we can synthesize the operator $\exp \left\{ i \gamma \hat{G} \right\} \hat{U}_{\rm PR}$ in a time that is infinitesimally close to $T$. It is worth noting here that a straightforward calculation based on counting the number of operators obtained from the commutation relations described above bounds $K$ between $\log_{M+1}[2(d^{2n}-1)/M]+1$ and $d^{2n}-M-1$.

While the above argument is conceptually clear from a mathematical point of view, because it separates the components $\exp \left\{ i \gamma \hat{G} \right\}$ and $\hat{U}_{\rm PR}$, a scenario that is closer to practical situations is one in which the control functions $f_j(t)$ are tweaked slightly to synthesize $\hat{U}_{\rm R}$ instead of $\hat{U}_{\rm PR}$. As can be seen from the inequality that we gave above for $T_{\rm R}$, the time needed to synthesize $\hat{U}_{\rm R}$ might in fact turn out to be shorter than the time needed to synthesize $\hat{U}_{\rm PR}$. In any case, the times needed to synthesize two unitary operators separated by an infinitesimally small distance will be equal, up to an infinitesimally small difference.

In the context of comparing the times needed to synthesize nearby unitary operators, it is worth mentioning the work of Ref.~\cite{Quillen}, which investigated chaotic dynamics in the context of synthesizing PRUs. The chaotic dynamics indicates that significantly different unitary operators can be synthesized using only slightly different control pulses. This result is consistent with our conclusion that a pair of unitary operators, one belonging to the RU ensemble and the other belonging to the PRU ensemble, can require only slightly different conditions to be synthesized in the continuous-time approach.

We should also point out that one can define an acceptable error $\epsilon$ and require that the prepared state or unitary operator is within distance $\epsilon$ of the target. In this case, any unitary operator that is within a distance $\epsilon$ of a PRU will require at most a polynomial amount of time to synthesize. One can then ask how much of the entire space is covered by PRSs or PRUs with a given value of $\epsilon$. A closely related question concerns the minimum value of $\epsilon$ that would allow the space of PRSs and PRUs to cover the entire space of states or unitary operators, respectively. Investigating these questions requires a dedicated study and goes beyond the scope of the present work. In this work, we assume that the acceptable error $\epsilon$ is effectively zero. Note that, to make sure that we are making a fair comparison between the computational and physical complexity, we apply the zero-error condition in both cases.

\subsection{Physical justification of setting numerical threshold for minimum gate time}

It is worth making a comment here about the fact that we set a certain value of the fidelity, e.g.~$F=0.999$, to identify the minimum gate time. In theory, for a fully controllable system with a finite-dimensional Hilbert space, it is possible to achieve perfect fidelity ($F=1$) in a finite time. More specifically, we expect $F$ to start off at a small value when the pulse time is zero, then increase following a smooth function until it reaches $F=1$ at some value of the pulse time and then remain at exactly $F=1$ after that. In our numerical calculations, the fidelity generally keeps increasing as we increase the evolution time. However, our experience has been that we sometimes encounter a near-stagnation where the rate of fidelity increase with increasing evolution time decreases significantly when $1-F\lesssim 10^{-3}$ for our numerical calculation settings, even in cases where there is no physical reason why the fidelity should stop increasing before it reaches $F=1$. We therefore believe that this near-stagnation is a result of slow convergence in the numerical optimization algorithm. Nevertheless, one can ask whether we are justified in using the threshold $F=0.999$ to identify the minimum synthesis time, or whether the time required to achieve $F=1$ could be significantly longer than the time needed to reach $F=0.999$.

From a physical point of view, if we obtain a fidelity of a few nines at a certain value of the evolution time, it means that the implemented unitary operator is very close to the target operator. We then need a unitary operator that deviates only slightly from the identity operator to transform the numerically obtained unitary operator to the target operator. Using an argument similar to that given in Section \ref{Sec:Discussion}.A, we can conclude that we need only a small amount of time to synthesize the final correction unitary operator and reach fidelity exactly equal to one. Setting a threshold fidelity of 0.999 for a few-qudit system with small qudit dimension $d$ is therefore a justified approach to accurately identify the minimum synthesis time.

\subsection{Physical complexity bounds based on parameter counting}

While there are no general physical complexity bounds for $n$ qudits similar to those derived in Sec.~\ref{Sec:ComputationalComplexity} for the computational complexity, there is a parameter counting procedure that could provide some insight into this problem.

First, we note that the physical complexity of two-qubit operations is well known \cite{Vidal}. The minimum time needed to synthesize an arbitrary two-qubit gate is proportional to $1/g$, where $g$ is the coupling strength between the two qubits \cite{Basyildiz2025}. Furthermore, similar bounds were recently derived for two-qubit gates when each logical qubit is obtained by using two quantum states of a $d$ dimensional physical system \cite{Basyildiz2023}. It should be noted, however, that these bounds are not tight for arbitrary two-qubit operations. 

To proceed with the parameter counting approach, we assume that we have a unitary operator $\hat{U}$, and we rewrite it in the form
\begin{equation}
\hat{U} = \left[ \bigotimes_{i=1}^n \hat{V}_i \right] e^{-i\hat{A}} \left[ \bigotimes_{i=1}^n \hat{W}_i \right],
\end{equation}
where $\hat{V}_i$ and $\hat{W}_i$ are $d$-dimensional rotations for the different qudits. In the case of two qubits, it is always possible to find $\hat{V}_i$ and $\hat{W}_i$ such that $\hat{A}$ takes the form
\begin{equation}
\hat{A} = c_1\hat{\sigma}_{x,1}\otimes\hat{\sigma}_{x,2} + c_2\hat{\sigma}_{y,1}\otimes\hat{\sigma}_{y,2} + c_3\hat{\sigma}_{z,1}\otimes\hat{\sigma}_{z,2},
\label{Eq:KrausDecomposition}
\end{equation}
where $\hat{\sigma}_{\alpha,i}$ (with $\alpha=x,y,z$ and $i=1,2$) are the different Pauli operators of the two qubits. We therefore have a convenient canonical decomposition of any unitary operator. The parameters $c_1$, $c_2$ and $c_3$ allow us to determine the minimum time required to synthesize $\hat{U}$ using the coupling Hamiltonian in the physical system under consideration \cite{Kraus,Vidal}.

Parameter counting can be introduced into this analysis as follows: a general 2-qubit unitary operator $\hat{U}$ has 15 free parameters. Each single-qubit rotation (among the four rotations $\hat{V}_i$ and $\hat{W}_i$) has 3 free parameters. Therefore, $\hat{A}$ must have at least $15-12=3$ parameters so that, when combined with $\hat{V}_i$ and $\hat{W}_i$, it will cover all possible unitary operators $\hat{U}$. As mentioned above, we can always find a decomposition with the three-parameter Hamiltonian in Eq.~(\ref{Eq:KrausDecomposition}). In other words, the Hamiltonian in Eq.~(\ref{Eq:KrausDecomposition}) contains the minimum of three parameters required to generate any unitary operator $\hat{U}$. Note that Eq.~(\ref{Eq:KrausDecomposition}) does not contain all the possible combinations of Pauli operators of two qubits, since there are nine such combinations in total.

We can generalize this exercise to $n$ qudits. The number of free parameters in $\hat{U}$ is $d^{2n}-1$. Each qudit rotation has $d^2-1$ free parameters. The number of free parameters in all of $\hat{V}_i$ and $\hat{W}_i$ combined is therefore $2n(d^2-1)$. Combining these numbers means that $\hat{A}$ must have $d^{2n} - 2nd^2 + 2n - 1$ free parameters. For example, for three qubits, we find that $\hat{A}$ must have 45 free parameters. There are only 16 different single-qubit and two-qubit Pauli operators. As a result, $\hat{A}$ cannot be expressed as the sum of only single-qubit and two-qubit terms. It must contain three-qubit terms.

In the two-qudit case, obviously any $\hat{A}$ can be written as a sum of single-qudit and two-qudit terms. The parameter counting procedure leads to the conclusion that $\hat{A}$ can have $n(d^2-1)$ fewer terms than the total number of possible combinations of two-qudit terms (which is given by $(d^2-1)^2$). We can therefore hope to find a form of $\hat{A}$ that allows us to derive a relationship between the inter-qudit coupling strength and the minimum gate time. Having said that, even for two qutrits, the number of free parameters in $\hat{A}$ is 48, which is much higher than the three parameters in Eq.~(\ref{Eq:KrausDecomposition}), making the derivation highly nontrivial. The generalization to more than two qudits leads to the conclusion that $\hat{A}$ cannot be expressed as the sum of only single-qudit and two-qudit terms.

To summarize, in both generalizations, i.e.~$n>2$ and/or $d>2$, the parameter counts quickly become large. Therefore, while it might be possible in principle, it is not at all straightforward to design and follow a procedure to decompose a general $n$-qudit unitary operator and use the decomposition to determine the physical complexity of synthesizing the operator.

\subsection{Alternative physical settings}

In this work, we considered a specific form of the Hamiltonian. As mentioned in Sec.~\ref{Sec:PhysicalComplexity}, we assumed a fully connected graph and fast single-qudit control. Furthermore, we used inter-qudit coupling terms that correspond to a bi-linear interaction between harmonic oscillators. This choice is motivated by experimental realizations of quantum computing devices, such as superconducting qudits. The qubit designs used in most present-day superconducting quantum processors are well described by models that treat them as weakly anharmonic oscillators. Related theoretical studies on synthesizing qudit operations in superconducting qudit systems were recently reported in Refs.~\cite{Fischer,Jaouadi}. It is important to keep in mind that our calculations can be modified to investigate any other physical realization of qudits, e.g.~coupling operators with matrix elements that do not increase as we go up the energy level ladder.

Similarly, we assumed that the different qudit transitions can be driven independently of each other. It is common to have situations where the applied control field couples to all transitions, and each transition responds mainly to the resonant component in the control field, while also being affected by near-resonant components. This situation generally leads to slower operations, especially for weakly anharmonic qudits \cite{Ashhab2022SpeedLimits,Basyildiz2025}.

\subsection{Computation time}

The calculations in this work took times ranging from one to a few hours per optimization. The variations in computation time include the expected increase in computation time with increasing system size, as well as variations between different (nominally similar) computers used to run the calculations.

We should also mention in this context that the computation time is not strongly correlated with the physical evolution time, i.e.~the total pulse time. In other words, calculating the optimal pulses required to perform the desired operation can take a long calculation time using classical OCT algorithms, even if the time needed to implement the desired quantum operation is small. One example of such a situation is obtained if we increase $d$ for $n=2$. As can be seen in Table \ref{Table:MinimumGateTime}, the minimum synthesis time decreases even though the Hilbert space size increases, which requires an increasing calculation time. A similar situation also occurs when we increase $n$, since the Hilbert space size grows exponentially with increasing $n$. One must therefore keep in mind that, in a practical setting, the classical computation time needed to calculate the control pulses should be included in any resource estimate for the total workflow.

\section{Conclusion}
\label{Sec:Conclusion}

In conclusion, we have analyzed the computational and physical complexity of synthesizing RSs and RUs of multi-qudit systems. We have derived mathematical lower bounds for the number of elementary gates needed to synthesize such states and unitary operators, and we have verified with numerical calculations that the lower bounds are quite tight. We have also performed OCT numerical calculations to identify the minimum time needed to synthesize RSs and RUs for few-qudit systems. While these small-scale calculations do not allow us to make definitive statements about the scaling laws for large systems, they provide some insight as to how the physical complexity changes when we vary the number of qudits and/or qudit dimension. We have discussed the relationship between RUs and PRUs and the difference in their synthesis complexity. Our results provide some guidelines for future work on controlling and utilizing multi-qudit systems.

\section*{Acknowledgments}

SA acknowledges support from Japan’s Ministry of Education, Culture, Sports, Science and Technology (MEXT) Quantum Leap Flagship Program Grant Number JPMXS0120319794 and the Center of Innovations for Sustainable Quantum AI (Grant No. JPMJPF2221) from the Japan Science and Technology Agency (JST). BB acknowledges funding provided by the Institute for Quantum Information and Matter, an NSF Physics Frontiers Center (NSF Grant PHY-2317110), and the JST Adopting Sustainable Partnerships for Innovative Research Ecosystem (ASPIRE) (Grant Number JPMJAP2513) through the RIKEN Center for Quantum Computing.


\begin{thebibliography}{99}
\bibitem{RandomUnitaryFootnote} This exponential scaling appears to run counter to the idea of speeding up computations using quantum computers. Indeed, it might seem surprising that implementing a unitary operator on a quantum computer requires an exponentially large number of gates. This situation raises fundamental questions about the role of RUs in quantum computation. We will not discuss these questions here.
\end{thebibliography}
\end{document}